\documentclass[preprint,amsmath,amssymb,aps,pra,longbibliography]{revtex4-1}

\usepackage[margin=0.75in, top=1in, bottom=0.85in]{geometry}

\usepackage[final,letterspace=-0]{microtype}
\usepackage[colorlinks=true, linkcolor=blue, urlcolor=blue, citecolor=blue, anchorcolor=blue]{hyperref}

\usepackage{mathpazo} 
\usepackage{upgreek}

\usepackage{graphicx}
\usepackage{dcolumn}
\usepackage{bm}
\usepackage{helvet}

\begin{document}

\title{Bridging the gap between ultrafast optics and resonant photonics via omni-resonance}

\author{Abbas Shiri$^{1}$, Kenneth L. Schepler$^{1}$, and Ayman F. Abouraddy$^{1}$}
\email{Corresponding author: raddy@creol.ucf.edu}
\affiliation{$^{1}$CREOL, The College of Optics \& Photonics, University of Central Florida, Orlando, FL 32816, USA}

\begin{abstract}
High-finesse planar Fabry-P{\'e}rot (FP) cavities spectrally filter the incident field at discrete resonances, and thus cannot be utilized to resonantly enhance the field of ultrashort pulses. Introducing judicious angular dispersion into a pulse can give rise to `omni-resonance', whereby the entire bandwidth of a spatiotemporally structured ultrafast pulse couples to a single longitudinal cavity resonance, even when the pulse bandwidth far exceeds the resonant linewidth. Here we show that omni-resonance increases the intra-cavity peak intensity above that of a pulse having equal energy and bandwidth when tightly focused in free space -- maintained across its entire bandwidth and along a cavity longer than the Rayleigh length of the focused pulse. This paves the way towards broadband resonant enhancement of nonlinear optical effects, thereby bridging the gap between ultrafast optics and resonant photonics.  
\end{abstract}

\maketitle

The spectral response of a planar Fabry-P{\'e}rot (FP) cavity is restricted to spectrally narrow linewidths at the resonant wavelengths \cite{Hernandez86Book,Vaughan89Book,Ismail16OE}. This renders short-length FP cavities inappropriate for harnessing field enhancement with ultrashort (broadband) pulses because of spectral filtering. Although the spectral narrowness of the resonances can be put to use in a variety of sensing applications \cite{Vahala03Nature}, it would be beneficial if resonantly enhanced interactions could be realized over a broad bandwidth; e.g., linear absorption in thin weakly absorbing layers via coherence perfect absorption \cite{Chong10PRL,Wan11S,Zhang12Light,Villinger15OL,Baranov17NRM,Pye17OL} would then become relevant for applications in photodetectors and solar cells. Efforts directed at modifying the cavity structure to broaden its resonant linewidth \textit{without} reducing the finesse include `white-light' cavities \cite{Wicht97OC} utilizing atomic \cite{Rinkleff05PS,Pati07PRL,Wu08PRA} or nonlinear \cite{Yum13JLT} resonances, among a host of other efforts \cite{Wise04CQG,Wise05PRL,Savchenkov06OL,Yum13OC,Kotlicki14OL}.

We have recently shown that spatiotemporally structuring a pulse \cite{Shen23JO,Abouraddy25OPN} by introducing angular dispersion \cite{Torres10AOP} to produce a space-time wave packet (STWP) \cite{Yessenov22AOP} can give rise to a configuration we refer to as `omni-resonance' \cite{Shabahang17SR}: a broadband ultrafast pulse that couples in its entirety -- without spectral filtering -- to a single longitudinal resonant mode, even if the pulse bandwidth far exceeds the resonant linewidth or even the cavity free spectral range (FSR) \cite{Shabahang19OL,Shiri20OL,Shiri20APLP,Shiri22OL,Hall25LPR}. By pre-conditioning the field without modifying the cavity, thus preserving its finesse, broadband resonant enhancement of linear absorption has been verified \cite{Villinger21AOM}. However, the spatiotemporal structuring required for omni-resonance delocalizes the pulse energy and thus reduces its peak intensity, which may diminish the utility of omni-resonance in enhancing nonlinear interactions such as two-photon absorption, Raman emission, and the Kerr effect. We pose here the following question: can the omni-resonant peak intensity exceed that of a conventional focused pulsed beam of equal energy and pulse width?

We answer this question in the affirmative by showing that omni-resonance can enhance the field within a short-length planar FP cavity whose FSR exceeds the pulse bandwidth, so that only a single resonance overlaps with the pulse spectrum. When a conventional ultrashort pulse is focused into a cavity, spectral filtering dominates at high finesse. In contrast, the entire spectrum of an omni-resonant STWP is coupled to the cavity (no spectral filtering). Critically, omni-resonance can enhance the cavity peak intensity above that of a tightly focused Gaussian pulse in free space of equal energy and spatial and temporal bandwidths -- maintained at high finesse  across the entire pulse bandwidth, along the cavity length (which can exceed the Rayleigh length of the free-space beam). One can thus benefit from resonantly enhancing the peak intensity of ultrashort laser pulses via omni-resonance for nonlinear effects, thereby bridging the gap between ultrafast optics and resonant photonics. 

\begin{figure*}[t!]
\centering
\includegraphics[width=17.4 cm]{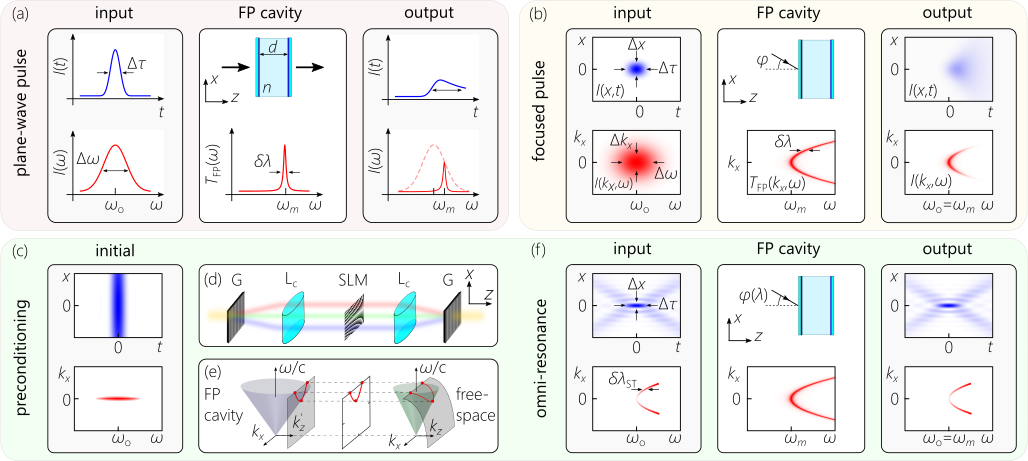}
\caption{(a) Coupling a plane-wave pulse to a planar FP cavity. The temporal profile and spectrum before and after the cavity are plotted on the left and right, respectively. In the middle we depict the FP cavity and the transfer function $T_{\mathrm{FP}}(0,\omega)$. (b) Coupling a focused Gaussian pulse to the FP cavity. The panels are similar to (a), but spatiotemporal profiles here replace the purely temporal counterparts. (c-f) Omni-resonance. (c) The spatiotemporal profile and spectrum for a plane-wave pulse. (d) Setup to pre-condition the plane-wave pulse for omni-resonance. (e) Spectral support for the omni-resonant field on the free-space and cavity light-cones, along with the spectral projection onto the $(k_{x},\tfrac{\omega}{c})$-plane. (f) Same as (b) for the omni-resonant configuration.}
\label{fig:Concept}
\end{figure*}

\textbf{FP cavity.} A planar FP cavity formed of a transparent layer of thickness $d$ and refractive index $n$ between symmetric mirrors of reflectivity $R$, with finesse $\mathcal{F}=\tfrac{\pi\sqrt{R}}{1-R}$ has a normalized spectral transmittance $T_{\mathrm{FP}}(k_{x},\omega)=1\big/\left(1+(\tfrac{2\mathcal{F}}{\pi})^{2}\sin^{2}\{\tfrac{\chi(k_{x},\omega)}{2}\}\right)$; where $k_{x}=\tfrac{2\pi}{\lambda}\sin\varphi$ is the transverse wave number along $x$, $\varphi$ is the external incident angle with the cavity normal, $\lambda$ is the free-space wavelength, $\omega$ is the temporal frequency, $\chi(k_{x},\omega)=\tfrac{4\pi nd}{\lambda}\sqrt{1-\tfrac{1}{n^{2}}\sin^{2}\varphi}$ is the cavity round-trip phase, and resonance occurs when $\chi=2\pi m$ (integer $m$ is the resonance order). The resonant wavelength is $\lambda_{m}(\varphi)=\lambda_{m}\sqrt{1-\tfrac{1}{n^{2}}\sin^{2}\varphi}$, where $\lambda_{m}=\tfrac{2nd}{m}$; at normal-incidence $\lambda_{m}(0^{\circ})=\lambda_{m}$, the FSR is $\approx\tfrac{\lambda_{m}}{m}$, the resonant linewidth is $\delta\lambda$, and increasing $\varphi$ (oblique incidence) blue-shifts the resonances \cite{Shabahang17SR}. The intensity of the cavity field is resonantly enhanced by $\sim\mathcal{F}$ with respect to the incident field at the resonant wavelengths coupled to the cavity. We take the FSR to be larger than the incident pulse bandwidth $\Delta\lambda$ (a single resonance overlaps with the pulse spectrum). We depict in Fig.~\ref{fig:Concept}(a) an example where $d=5$~$\mu$m, $n=1.5$, and $\mathcal{F}=61$, $\lambda_{m}=1$~$\mu$m ($m=15$), $\mathrm{FSR}\approx62.5$~nm, and resonant linewidth is $\delta\lambda=1.1$~nm. A plane-wave pulse of width $\Delta\tau\approx220$~fs with $\Delta\lambda\approx6.7$~nm normally incident on the cavity is temporally broadened to $\sim600$~fs through spectral filtering.

We write the field as $E(x,z;t)=e^{i(k_{\mathrm{o}}z-\omega_{\mathrm{o}}t)}\psi(x,z;t)$, where $\psi(x,z;t)\!=\!\iint\!dk_{x}d\Omega\widetilde{\psi}(k_{x},\Omega)e^{i\{k_{x}x+(k_{z}-k_{\mathrm{o}})z-\Omega t\}}$ is the envelope, $\omega_{\mathrm{o}}$ is the carrier (temporal) frequency, $k_{\mathrm{o}}=\omega_{\mathrm{o}}/c$, $c$ is the speed of light in vacuum, $\Omega=\omega-\omega_{\mathrm{o}}$, $k_{z}=\tfrac{\omega}{c}\cos\varphi$ is the axial wave number along $z$, respectively, and $\varphi$ is the propagation angle with the $z$-axis (normal to the FP cavity). It will suffice to consider only one transverse spatial dimension $x$ to achieve omni-resonance. Because $k_{x}$ and $\Omega$ are invariant across a planar interface between two dielectric media, the spatiotemporal spectrum $\widetilde{\psi}(k_{x},\Omega)$ is the same in free space and inside the cavity \cite{Bhaduri20NatPhot}. We compare here two field configurations in which we hold \textit{fixed} the energy $\mathcal{E}=\iint\!dxdt|E(x,z;t)|^{2}$ and the temporal bandwidth $\Delta\lambda$: a focused Gaussian pulse for conventional resonance [Fig.~\ref{fig:Concept}(b)] and an STWP for omni-resonance [Fig.~\ref{fig:Concept}(c-f)].

\textbf{Focused Gaussian pulse.} Consider first a focused Gaussian pulse  [Fig.~\ref{fig:Concept}(b)], $\psi(x,0;t)\propto\exp\left\{-\tfrac{x^{2}}{2(\Delta x)^{2}}\right\}\exp\left\{-\tfrac{t^{2}}{2(\Delta\tau)^{2}}\right\}$, where $\Delta x$ and $\Delta\tau$ are the spatial and temporal widths, respectively [Fig.~\ref{fig:Concept}(b), left]. The corresponding spatial and temporal bandwidths $\Delta k_{x}$ and $\Delta\omega$ are -- in principle -- independent of each other. Focusing this pulse (reducing $\Delta x$) increases the pulse peak intensity, which is crucial for nonlinear optics, but this comes at the price of reducing the propagation length over which it is maintained \cite{Benabid02Science}. The field coupled to the cavity is determined by the overlap between the spatiotemporal spectrum $|\widetilde{\psi}(k_{x},\omega)|^{2}$ and the spectral transmission $T_{\mathrm{FP}}(k_{x},\omega)$ for the $m^{\mathrm{th}}$ resonance [Fig.~\ref{fig:Concept}(b), center], with $\omega_{\mathrm{o}}=\omega_{m}$ for optimal coupling. Only a fraction of the pulse energy is coupled to the cavity because of spatiotemporal filtering [Fig.~\ref{fig:Concept}(b), right], so the transmitted spatiotemporal field profile departs from that of the incident. 

\textbf{Omni-resonant STWP.} Consider next an STWP [Fig.~\ref{fig:Concept}(c-f)] in which each $k_{x}$ is associated with a single $\omega$ \cite{Kondakci17NP,Yessenov22AOP}, obtained by preconditioning a plane-wave pulse [Fig.~\ref{fig:Concept}(c)] via the setup in Fig.~\ref{fig:Concept}(d). A diffraction grating (G) and cylindrical lens (L$_{\mathrm{c}}$) spatially resolve the temporal spectrum, and a spatial light modulator (SLM) imparts a prescribed $\pm k_{x}(\lambda)$ to each $\lambda$ \cite{Hall24OE}, to within a spectral uncertainty $\delta\lambda_{\mathrm{ST}}$, such that the cavity axial wave number $k_{z}'=k_{m}=n\tfrac{2\pi}{\lambda_{m}}$ is constant [Fig.~\ref{fig:Concept}(e)], thereby satisfying the resonant condition $\chi(k_{x},\lambda)=2\pi m$ for fixed $m$ across the entire pulse bandwidth \cite{Hall25LPR}. In the vicinity of $k_{x}=0$ this requires restricting the spectral support for the omni-resonant STWP on the cavity light-cone $k_{x}^{2}+k_{z}'^{2}=(n\tfrac{\omega}{c})^{2}$ to a parabola $\tfrac{\Omega(k_{x})}{\omega_{\mathrm{o}}}\approx\tfrac{k_{x}^{2}}{2n_{\mathrm{o}}^{2}k_{\mathrm{o}}^{2}}$ \cite{Bhaduri20NatPhot}, which corresponds to the intersection of the cavity light-cone surface with a vertical plane $k_{z}'=k_{m}$. The parabola projected onto the $(k_{x},\tfrac{\omega}{c})$-plane is invariant across planar interfaces \cite{Bhaduri20NatPhot}, which is the basis for synthesizing the field in free space via the setup in Fig.~\ref{fig:Concept}(d) \cite{Shiri20OL}; see the spectral support on the free-space light-cone $k_{x}^{2}+k_{z}^{2}=(\tfrac{\omega}{c})^{2}$ in Fig.~\ref{fig:Concept}(e). The tight association between $k_{x}$ and $\omega$ enforces a relationship between $\Delta k_{x}$ and $\Delta\omega$: $\tfrac{\Delta\omega}{\omega_{\mathrm{o}}}=\tfrac{1}{2n^{2}}(\tfrac{\Delta k_{x}}{k_{\mathrm{o}}})^{2}$; here $\Delta x=1.3$~$\mu$m at $\Delta\lambda=6.7$~nm. The spatiotemporal intensity profile for this STWP is X-shaped \cite{Yessenov22AOP} and its spectrum $|\widetilde{\psi}(k_{x},\Omega)|^{2}$ \textit{matches} the cavity spectral transmission $T_{\mathrm{FP}}(k_{x},\omega)$ [Fig.~\ref{fig:Concept}(f)]. Therefore, when $\delta\lambda_{\mathrm{ST}}<\delta\lambda$, the entire energy of the STWP couples to the cavity and resonates within it. As such, the spatiotemporal profile and spectrum at the cavity output match those at the input [Fig.~\ref{fig:Concept}(f)].

\begin{figure}[t!]
\centering
\includegraphics[width=8.6cm]{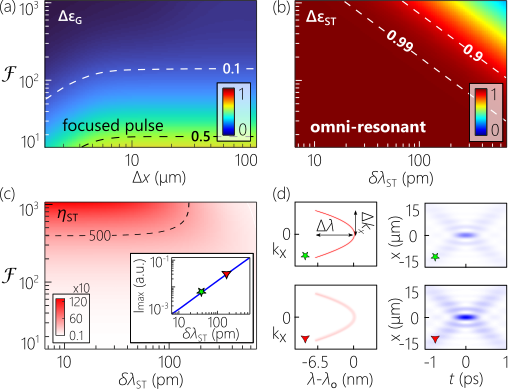}
\caption{(a) Energy fraction coupled to the cavity from a focused Gaussian pulse $\Delta\mathcal{E}_{\mathrm{G}}$ and (b) an omni-resonant STWP $\Delta\mathcal{E}_{\mathrm{ST}}$. (c) Cavity enhancement $\eta_{\mathrm{ST}}$ for the omni-resonant STWP with respect to free space. The inset shows the STWP free-space peak intensity while varying $\delta\lambda_{\mathrm{ST}}$. (d) The spatiotemporal spectrum $|\widetilde{\psi}(k_{x},\omega)|^{2}$ and intensity profile $I(x,z=0;t)$ for two omni-resonant STWPs with $\delta\lambda_{\mathrm{ST}}=40$~pm and 200~pm, corresponding to the points in the inset in (c).}
\label{fig:GaussianVsSTWP}
\end{figure}

\textbf{Coupling efficiency.} We plot in Fig.~\ref{fig:GaussianVsSTWP} the calculated fraction of the pulse energy $\Delta\mathcal{E}=\mathcal{E}_{\mathrm{out}}/\mathcal{E}_{\mathrm{in}}$ coupled to the cavity for the two cases considered ($\mathcal{E}_{\mathrm{in}}$ and $\mathcal{E}_{\mathrm{out}}$ are the incident and transmitted energies, respectively), which is determined by the overlap between $|\widetilde{\psi}(k_{x},\omega)|^{2}$ and $T_{\mathrm{FP}}(k_{x},\omega)$ for $d=5$~$\upmu$m and $\Delta\lambda=6.7$~nm. We plot $\Delta\mathcal{E}_{\mathrm{G}}$ for the focused Gaussian pulse in Fig.~\ref{fig:GaussianVsSTWP}(a) as we vary the focused beam size $\Delta x$ (which is independent of $\Delta\lambda$) and the cavity finesse $\mathcal{F}$. In this scenario, $\Delta\mathcal{E}_{\mathrm{G}}$ drops rapidly with increasing $\mathcal{F}$ (due to spectral filtering), and $\Delta\mathcal{E}_{\mathrm{G}}$ is independent of $\Delta x$ when $\Delta x$ is large, but $\Delta\mathcal{E}_{\mathrm{G}}$ drops slightly at tight focusing (small $\Delta x$) because the concomitant increase in $\Delta k_{x}$ reduces the overlap between $|\widetilde{\psi}(k_{x},\omega)|^{2}$ and $T_{\mathrm{FP}}(k_{x},\omega)$ [Fig.~\ref{fig:Concept}(b)]. For $\mathcal{F}\sim100$, only $10\%$ of this pulse energy is coupled to the cavity. Changing the cavity length $d$ or the bandwidth $\Delta\lambda$ modifies $\Delta\mathcal{E}_{\mathrm{G}}$: $\Delta\mathcal{E}_{\mathrm{G}}$ drops with larger $d$ because of the associated narrowing in the resonant linewidth $\delta\lambda$, and $\Delta\mathcal{E}_{\mathrm{G}}$ also drops with larger $\Delta\lambda$ because of the more significant impact of spectral filtering (Supplementary Fig.~S1). We plot $\Delta\mathcal{E}_{\mathrm{ST}}$ for an omni-resonant STWP in Fig.~\ref{fig:GaussianVsSTWP}(b) as we vary the spectral uncertainty $\delta\lambda_{\mathrm{ST}}$ and $\mathcal{F}$. Holding $\Delta\lambda=6.7$~nm fixed as in Fig.~\ref{fig:GaussianVsSTWP}(a) in turn fixes $\Delta x\approx1.3$~$\upmu$m. We find that $\Delta\mathcal{E}_{\mathrm{ST}}>0.99$ across most of this parameter space, and $\Delta\mathcal{E}_{\mathrm{ST}}$ drops only for high $\mathcal{F}$ and large $\delta\lambda_{\mathrm{ST}}$ whereupon $\delta\lambda_{\mathrm{ST}}>\delta\lambda$ (resulting in spectral filtering). Changing $d$ or $\Delta\lambda$ modifies this coupling efficiency only slightly (Supplementary Fig.~S2).

\textbf{Resonant enhancement.} Because the omni-resonant STWP energy is coupled in its entirety to the cavity without spectral filtering (as long as $\delta\lambda_{\mathrm{ST}}<\delta\lambda$), the peak cavity intensity $I_{\mathrm{peak}}^{\mathrm{ST,FP}}$ is enhanced with respect to the incident free-space STWP peak intensity $I_{\mathrm{peak}}^{\mathrm{ST,free}}$ by the factor $\eta_{\mathrm{ST}}=I_{\mathrm{peak}}^{\mathrm{ST,FP}}/I_{\mathrm{peak}}^{\mathrm{ST,free}}\sim\mathcal{F}$ [Fig.~\ref{fig:GaussianVsSTWP}(c)]. However, to ascertain whether omni-resonance can facilitate broadband resonance enhancement of nonlinear effects, we must compare $I_{\mathrm{peak}}^{\mathrm{ST,FP}}$ to that for a focused Gaussian pulse of the same $\mathcal{E}$, $\Delta\lambda$, and $\Delta k_{x}$. Focusing a conventional pulse in space and time increases $I_{\mathrm{peak}}^{\mathrm{G,free}}$, but this intensity is maintained over an increasingly shorter distance (the Rayleigh length), whereas $I_{\mathrm{peak}}^{\mathrm{ST,FP}}$  is maintained along the cavity length. However, the X-shaped profile for the STWP makes it less spatially localized than a focused Gaussian pulse. Although this spatial localization is enhanced ($I_{\mathrm{peak}}^{\mathrm{ST,free}}$ increased) by increasing $\delta\lambda_{\mathrm{ST}}$ [Fig.~\ref{fig:GaussianVsSTWP}(c) inset and Fig.~\ref{fig:GaussianVsSTWP}(d)], what is gained in enhancing $I_{\mathrm{peak}}^{\mathrm{ST,free}}$ by increasing $\delta\lambda_{\mathrm{ST}}$ is offset once the STWP is coupled due to the decreased $\Delta\mathcal{E}_{\mathrm{ST}}$ at high $\mathcal{F}$, where $\delta\lambda_{\mathrm{ST}}>\delta\lambda$ results in spectral filtering.

\begin{figure}[t!]
\centering
\includegraphics[width=8.6cm]{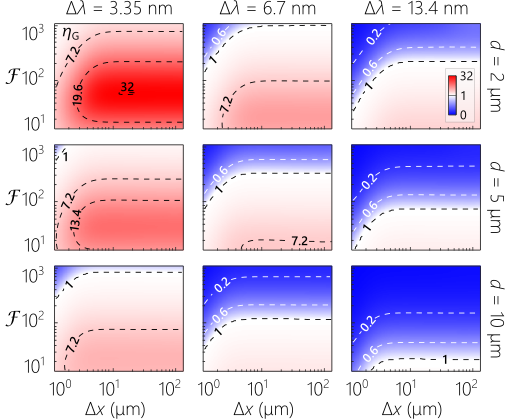}
\caption{Field resonant enhancement $\eta_{\mathrm{G}}$ for a focused Gaussian pulse with respect to its free-space counterpart. The columns correspond to different bandwidths $\Delta\lambda$, and the rows to different cavity lengths $d$. Two color palettes distinguish the regimes $\eta_{\mathrm{G}}<1$ (blue) and $\eta_{\mathrm{G}}>1$ (red).}
\label{fig:Gaussian}
\end{figure}

\begin{figure}[t!]
\centering
\includegraphics[width=8.6cm]{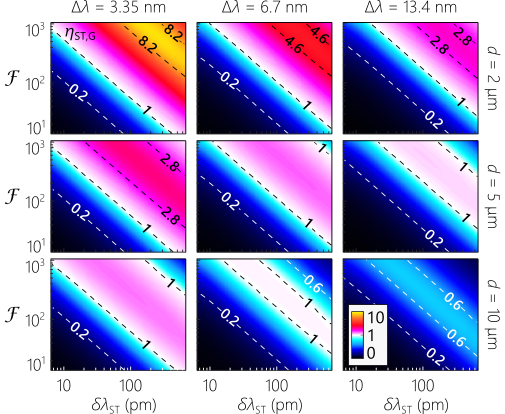}
\caption{Cavity enhancement $\eta_{\mathrm{ST,G}}$ for an omni-resonant STWP with respect to a free-space focused Gaussian pulse (holding $\mathcal{E}$ and $\Delta x$ fixed). The columns correspond to bandwidths $\Delta\lambda$, and the rows to cavity lengths $d$. Two color palettes distinguish the regimes $\eta_{\mathrm{ST,G}}<1$ (blue) and $\eta_{\mathrm{ST,G}}>1$ (red).}
\label{fig:Omniresonant}
\end{figure}

For reference, we first plot in Fig.~\ref{fig:Gaussian} the enhancement factor $\eta_{\mathrm{G}}=I_{\mathrm{peak}}^{\mathrm{G,FP}}/I_{\mathrm{peak}}^{\mathrm{G,free}}$ for three focused Gaussian pulses of bandwidth $\Delta\lambda=3.35,6.7$, and 13.4~nm (pulse widths of 440, 220, and 110~fs, respectively, at $\lambda_{\mathrm{o}}=1$~$\upmu$m) in FP cavities of length $d=2,5$, and 10~$\upmu$m. In each panel in Fig.~\ref{fig:Gaussian} (corresponding to a particular combination of $\Delta\lambda$ and $d$) we vary $\mathcal{F}$ and $\Delta x$. The FP cavity enhances the field when both $\Delta\lambda$ and $d$ are small, whereupon $\delta\lambda$ increases, thereby reducing spectral filtering, so that the resonant enhancement can counterbalance the drop in coupled energy $\Delta\mathcal{E}_{\mathrm{G}}$. Therefore, $\eta_{\mathrm{G}}>1$ here is still associated with spectral filtering and pulse broadening. In the opposite regime of large $\Delta\lambda$ (ultrashort pulse) and large $d$ (reduced $\delta\lambda$), the intra-cavity field is diminished with respect to its free-space counterpart: the field is heavily filtered (the pulse width increases), and the reduction in $\Delta\mathcal{E}_{\mathrm{G}}$ (Supplementary Fig.~S1) cannot be counterbalanced by the resonant field enhancement. Consequently, such planar FP cavities cannot resonantly enhance ultrashort pulses.
We plot in Fig.~\ref{fig:Omniresonant} the enhancement factor $\eta_{\mathrm{ST,G}}=I_{\mathrm{peak}}^{\mathrm{ST,FP}}/I_{\mathrm{peak}}^{\mathrm{G,free}}$, where we compare the resonantly enhanced intra-cavity peak intensity for the STWP to the peak intensity for the focused Gaussian pulse in free space. We use the same values of $d$ and $\Delta\lambda$ from Fig.~\ref{fig:Gaussian}. Because $\Delta\lambda$ and $\Delta k_{x}$ are related, the bandwidths $\Delta\lambda$ selected correspond to beam sizes $\Delta x=1.8$,1.3, and 0.9~$\upmu$m, respectively. In each panel in Fig.~\ref{fig:Omniresonant} we vary $\delta\lambda_{\mathrm{ST}}$ and $\mathcal{F}$, and use the focused Gaussian pulse with equal $\Delta x$ for normalization. We find that $\eta_{\mathrm{ST,G}}>1$ over a broad range of the parameters $\mathcal{F}$ and $\delta\lambda_{\mathrm{ST}}$. However, even when $\eta_{\mathrm{ST,G}}\sim1$, omni-resonance still offers an advantage. The intra-cavity intensity matches that of a tightly focused conventional pulse of the same energy and bandwidth \textit{without spectral filtering} (no pulse broadening; Supplementary Fig.~S2). Moreover, this intra-cavity peak is maintained along the cavity, which exceeds the Rayleigh length of the tightly focused pulse used as reference, thereby offering a longer interaction length for the ultrashort pulse. In other words, omni-resonance provides resonant field enhancement over a broad bandwidth along the cavity, while producing higher peak intensities than a focused pulse in free space.

\textbf{Discussion.} The concept of omni-resonance can be extended in several directions. First, the omni-resonance bandwidth can be dramatically increased \cite{Hall25LPR}. The maximum bandwidth that can be associated with a single resonance is $\Delta\lambda_{m}=\lambda_{m}(0^{\circ})-\lambda_{m}(90^{\circ})$. With an external angular acceptance $\varphi_{\mathrm{max}}=45^{\circ}$ we have $\Delta\lambda_{m}\approx118$~nm (pulse width $\approx 10$~fs at $\lambda_{\mathrm{o}}=1$~$\upmu$m), and $\Delta\lambda_{m}=15$~nm for $\varphi_{\mathrm{max}}=10^{\circ}$. The omni-resonance bandwidth can be increased at oblique incidence on the FP cavity; e.g., tilting the cavity by $45^{\circ}$ yields $\Delta\lambda_{m}\approx40$~nm (pulse width $\approx 26$~fs at $\lambda_{\mathrm{o}}=1$~$\upmu$m) for $\varphi_{\mathrm{max}}=10^{\circ}$. Second, the results can be generalized to two transverse dimensions \cite{Yessenov22NC,Yessenov25JOSAA,Yessenov25NC} and adapted to dispersive cavities \cite{Hall23LPR}. Third, ultra-compact STWP synthesizer can be made via a new class of rotated chirped Bragg volumetric gratings \cite{Yessenov23OL}. Finally, more work is needed for reduced FSR, where the spectrum of a conventional pulse couples to multiple longitudinal resonances, while the omni-resonant field remains associated with a single resonance \cite{Hall25LPR}.

\textbf{Conclusion.} In conclusion, we have shown that omni-resonance in a planar FP cavity can produce broadband resonant enhancement of the intra-cavity peak intensity compared to that of a tightly focused conventional pulse in free space having the same energy and bandwidth. The pulse suffers no spectral filtering even when its spectrum overlaps with a single narrow resonance, and the enhanced intensity is maintained along a cavity longer than the Rayleigh length of the focused pulse. Omni-resonance can thus help bridge the gap between ultrafast optics and resonant photonics.

\vspace{2mm}
\noindent
\textbf{Acknowledgments}\\
U.S. Office of Naval Research (ONR) N00014-20-1-2789.

\bibliography{diffraction}

\begin{thebibliography}{40}%
\makeatletter
\providecommand \@ifxundefined [1]{%
 \@ifx{#1\undefined}
}%
\providecommand \@ifnum [1]{%
 \ifnum #1\expandafter \@firstoftwo
 \else \expandafter \@secondoftwo
 \fi
}%
\providecommand \@ifx [1]{%
 \ifx #1\expandafter \@firstoftwo
 \else \expandafter \@secondoftwo
 \fi
}%
\providecommand \natexlab [1]{#1}%
\providecommand \enquote  [1]{``#1''}%
\providecommand \bibnamefont  [1]{#1}%
\providecommand \bibfnamefont [1]{#1}%
\providecommand \citenamefont [1]{#1}%
\providecommand \href@noop [0]{\@secondoftwo}%
\providecommand \href [0]{\begingroup \@sanitize@url \@href}%
\providecommand \@href[1]{\@@startlink{#1}\@@href}%
\providecommand \@@href[1]{\endgroup#1\@@endlink}%
\providecommand \@sanitize@url [0]{\catcode `\\12\catcode `\$12\catcode `\&12\catcode `\#12\catcode `\^12\catcode `\_12\catcode `\%12\relax}%
\providecommand \@@startlink[1]{}%
\providecommand \@@endlink[0]{}%
\providecommand \url  [0]{\begingroup\@sanitize@url \@url }%
\providecommand \@url [1]{\endgroup\@href {#1}{\urlprefix }}%
\providecommand \urlprefix  [0]{URL }%
\providecommand \Eprint [0]{\href }%
\providecommand \doibase [0]{http://dx.doi.org/}%
\providecommand \selectlanguage [0]{\@gobble}%
\providecommand \bibinfo  [0]{\@secondoftwo}%
\providecommand \bibfield  [0]{\@secondoftwo}%
\providecommand \translation [1]{[#1]}%
\providecommand \BibitemOpen [0]{}%
\providecommand \bibitemStop [0]{}%
\providecommand \bibitemNoStop [0]{.\EOS\space}%
\providecommand \EOS [0]{\spacefactor3000\relax}%
\providecommand \BibitemShut  [1]{\csname bibitem#1\endcsname}%
\let\auto@bib@innerbib\@empty
\bibitem [{\citenamefont {Hernandez}(1986)}]{Hernandez86Book}%
  \BibitemOpen
  \bibfield  {author} {\bibinfo {author} {\bibfnamefont {G.}~\bibnamefont {Hernandez}},\ }\href@noop {} {\emph {\bibinfo {title} {Fabry-Perot Interferometers}}}\ (\bibinfo  {publisher} {Cambridge Univ. Press},\ \bibinfo {address} {Cambridge},\ \bibinfo {year} {1986})\BibitemShut {NoStop}%
\bibitem [{\citenamefont {Vaughan}(1989)}]{Vaughan89Book}%
  \BibitemOpen
  \bibfield  {author} {\bibinfo {author} {\bibfnamefont {J.~M.}\ \bibnamefont {Vaughan}},\ }\href@noop {} {\emph {\bibinfo {title} {The Fabry-Perot Interferometer}}}\ (\bibinfo  {publisher} {Adam Hilger},\ \bibinfo {address} {Bristol},\ \bibinfo {year} {1989})\BibitemShut {NoStop}%
\bibitem [{\citenamefont {Ismail}\ \emph {et~al.}(2016)\citenamefont {Ismail}, \citenamefont {Kores}, \citenamefont {Geskus},\ and\ \citenamefont {Pollnau}}]{Ismail16OE}%
  \BibitemOpen
  \bibfield  {author} {\bibinfo {author} {\bibfnamefont {N.}~\bibnamefont {Ismail}}, \bibinfo {author} {\bibfnamefont {C.~C.}\ \bibnamefont {Kores}}, \bibinfo {author} {\bibfnamefont {D.}~\bibnamefont {Geskus}}, \ and\ \bibinfo {author} {\bibfnamefont {M.}~\bibnamefont {Pollnau}},\ }\bibfield  {title} {\enquote {\bibinfo {title} {Fabry-{P\'e}rot resonator: spectral line shapes, generic and related {A}iry distributions, linewidths, finesses, and performance at low or frequency-dependent reflectivity},}\ }\href@noop {} {\bibfield  {journal} {\bibinfo  {journal} {Opt. Express}\ }\textbf {\bibinfo {volume} {24}},\ \bibinfo {pages} {16366--16389} (\bibinfo {year} {2016})}\BibitemShut {NoStop}%
\bibitem [{\citenamefont {Vahala}(2003)}]{Vahala03Nature}%
  \BibitemOpen
  \bibfield  {author} {\bibinfo {author} {\bibfnamefont {K.~J.}\ \bibnamefont {Vahala}},\ }\bibfield  {title} {\enquote {\bibinfo {title} {Optical microcavities},}\ }\href@noop {} {\bibfield  {journal} {\bibinfo  {journal} {Nature}\ }\textbf {\bibinfo {volume} {424}},\ \bibinfo {pages} {839--846} (\bibinfo {year} {2003})}\BibitemShut {NoStop}%
\bibitem [{\citenamefont {Chong}\ \emph {et~al.}(2010)\citenamefont {Chong}, \citenamefont {Ge}, \citenamefont {Cao},\ and\ \citenamefont {Stone}}]{Chong10PRL}%
  \BibitemOpen
  \bibfield  {author} {\bibinfo {author} {\bibfnamefont {Y.~D.}\ \bibnamefont {Chong}}, \bibinfo {author} {\bibfnamefont {L.}~\bibnamefont {Ge}}, \bibinfo {author} {\bibfnamefont {H.}~\bibnamefont {Cao}}, \ and\ \bibinfo {author} {\bibfnamefont {A.~D.}\ \bibnamefont {Stone}},\ }\bibfield  {title} {\enquote {\bibinfo {title} {Coherent perfect absorbers: Time-reversed lasers},}\ }\href@noop {} {\bibfield  {journal} {\bibinfo  {journal} {Phys. Rev. Lett.}\ }\textbf {\bibinfo {volume} {105}},\ \bibinfo {pages} {053901} (\bibinfo {year} {2010})}\BibitemShut {NoStop}%
\bibitem [{\citenamefont {Wan}\ \emph {et~al.}(2011)\citenamefont {Wan}, \citenamefont {Chong}, \citenamefont {Ge}, \citenamefont {Noh}, \citenamefont {Stone},\ and\ \citenamefont {Cao}}]{Wan11S}%
  \BibitemOpen
  \bibfield  {author} {\bibinfo {author} {\bibfnamefont {W.}~\bibnamefont {Wan}}, \bibinfo {author} {\bibfnamefont {Y.}~\bibnamefont {Chong}}, \bibinfo {author} {\bibfnamefont {L.}~\bibnamefont {Ge}}, \bibinfo {author} {\bibfnamefont {H.}~\bibnamefont {Noh}}, \bibinfo {author} {\bibfnamefont {A.~D.}\ \bibnamefont {Stone}}, \ and\ \bibinfo {author} {\bibfnamefont {H.}~\bibnamefont {Cao}},\ }\bibfield  {title} {\enquote {\bibinfo {title} {Time-reversed lasing and interferometric control of absorption},}\ }\href@noop {} {\bibfield  {journal} {\bibinfo  {journal} {Science}\ }\textbf {\bibinfo {volume} {331}},\ \bibinfo {pages} {889--892} (\bibinfo {year} {2011})}\BibitemShut {NoStop}%
\bibitem [{\citenamefont {Zhang}\ \emph {et~al.}(2012)\citenamefont {Zhang}, \citenamefont {MacDonald},\ and\ \citenamefont {Zheludev}}]{Zhang12Light}%
  \BibitemOpen
  \bibfield  {author} {\bibinfo {author} {\bibfnamefont {J.}~\bibnamefont {Zhang}}, \bibinfo {author} {\bibfnamefont {K.~F.}\ \bibnamefont {MacDonald}}, \ and\ \bibinfo {author} {\bibfnamefont {N.~I.}\ \bibnamefont {Zheludev}},\ }\bibfield  {title} {\enquote {\bibinfo {title} {Controlling light-with-light without nonlinearity},}\ }\href@noop {} {\bibfield  {journal} {\bibinfo  {journal} {Light Sci. \& Appl.}\ }\textbf {\bibinfo {volume} {1}},\ \bibinfo {pages} {e18} (\bibinfo {year} {2012})}\BibitemShut {NoStop}%
\bibitem [{\citenamefont {Villinger}\ \emph {et~al.}(2015)\citenamefont {Villinger}, \citenamefont {Bayat}, \citenamefont {Pye},\ and\ \citenamefont {Abouraddy}}]{Villinger15OL}%
  \BibitemOpen
  \bibfield  {author} {\bibinfo {author} {\bibfnamefont {M.~L.}\ \bibnamefont {Villinger}}, \bibinfo {author} {\bibfnamefont {M.}~\bibnamefont {Bayat}}, \bibinfo {author} {\bibfnamefont {L.~N.}\ \bibnamefont {Pye}}, \ and\ \bibinfo {author} {\bibfnamefont {A.~F.}\ \bibnamefont {Abouraddy}},\ }\bibfield  {title} {\enquote {\bibinfo {title} {Analytical model for coherent perfect absorption in one-dimensional photonic structures},}\ }\href@noop {} {\bibfield  {journal} {\bibinfo  {journal} {Opt. Lett.}\ }\textbf {\bibinfo {volume} {40}},\ \bibinfo {pages} {5550--5553} (\bibinfo {year} {2015})}\BibitemShut {NoStop}%
\bibitem [{\citenamefont {Baranov}\ \emph {et~al.}(2017)\citenamefont {Baranov}, \citenamefont {Krasnok}, \citenamefont {Shegai}, \citenamefont {Al{\'u}},\ and\ \citenamefont {Chong}}]{Baranov17NRM}%
  \BibitemOpen
  \bibfield  {author} {\bibinfo {author} {\bibfnamefont {D.~G.}\ \bibnamefont {Baranov}}, \bibinfo {author} {\bibfnamefont {A.}~\bibnamefont {Krasnok}}, \bibinfo {author} {\bibfnamefont {T.}~\bibnamefont {Shegai}}, \bibinfo {author} {\bibfnamefont {A.}~\bibnamefont {Al{\'u}}}, \ and\ \bibinfo {author} {\bibfnamefont {Y.}~\bibnamefont {Chong}},\ }\bibfield  {title} {\enquote {\bibinfo {title} {Coherent perfect absorbers: linear control of light with light},}\ }\href@noop {} {\bibfield  {journal} {\bibinfo  {journal} {Nat. Rev. Mater.}\ }\textbf {\bibinfo {volume} {2}},\ \bibinfo {pages} {17064} (\bibinfo {year} {2017})}\BibitemShut {NoStop}%
\bibitem [{\citenamefont {Pye}\ \emph {et~al.}(2017)\citenamefont {Pye}, \citenamefont {Villinger}, \citenamefont {Shabahang}, \citenamefont {Larson}, \citenamefont {Martin},\ and\ \citenamefont {Abouraddy}}]{Pye17OL}%
  \BibitemOpen
  \bibfield  {author} {\bibinfo {author} {\bibfnamefont {L.~N.}\ \bibnamefont {Pye}}, \bibinfo {author} {\bibfnamefont {M.~L.}\ \bibnamefont {Villinger}}, \bibinfo {author} {\bibfnamefont {S.}~\bibnamefont {Shabahang}}, \bibinfo {author} {\bibfnamefont {W.~D.}\ \bibnamefont {Larson}}, \bibinfo {author} {\bibfnamefont {L.}~\bibnamefont {Martin}}, \ and\ \bibinfo {author} {\bibfnamefont {A.~F.}\ \bibnamefont {Abouraddy}},\ }\bibfield  {title} {\enquote {\bibinfo {title} {Octave-spanning coherent perfect absorption in a thin silicon film},}\ }\href@noop {} {\bibfield  {journal} {\bibinfo  {journal} {Opt. Lett.}\ }\textbf {\bibinfo {volume} {42}},\ \bibinfo {pages} {151--154} (\bibinfo {year} {2017})}\BibitemShut {NoStop}%
\bibitem [{\citenamefont {Wicht}\ \emph {et~al.}(1997)\citenamefont {Wicht}, \citenamefont {Danzmann}, \citenamefont {Fleischhauer}, \citenamefont {Scully}, \citenamefont {M{\"u}ller},\ and\ \citenamefont {Rinkleff}}]{Wicht97OC}%
  \BibitemOpen
  \bibfield  {author} {\bibinfo {author} {\bibfnamefont {A.}~\bibnamefont {Wicht}}, \bibinfo {author} {\bibfnamefont {K.}~\bibnamefont {Danzmann}}, \bibinfo {author} {\bibfnamefont {M.}~\bibnamefont {Fleischhauer}}, \bibinfo {author} {\bibfnamefont {M.}~\bibnamefont {Scully}}, \bibinfo {author} {\bibfnamefont {G.}~\bibnamefont {M{\"u}ller}}, \ and\ \bibinfo {author} {\bibfnamefont {R.-H.}\ \bibnamefont {Rinkleff}},\ }\bibfield  {title} {\enquote {\bibinfo {title} {White-light cavities, atomic phase coherence, and gravitational wave detectors},}\ }\href@noop {} {\bibfield  {journal} {\bibinfo  {journal} {Opt. Commun.}\ }\textbf {\bibinfo {volume} {134}},\ \bibinfo {pages} {431--439} (\bibinfo {year} {1997})}\BibitemShut {NoStop}%
\bibitem [{\citenamefont {Rinkleff}\ and\ \citenamefont {Wicht}(2005)}]{Rinkleff05PS}%
  \BibitemOpen
  \bibfield  {author} {\bibinfo {author} {\bibfnamefont {R.-H.}\ \bibnamefont {Rinkleff}}\ and\ \bibinfo {author} {\bibfnamefont {A}~\bibnamefont {Wicht}},\ }\bibfield  {title} {\enquote {\bibinfo {title} {The concept of white light cavities using atomic phase coherence},}\ }\href@noop {} {\bibfield  {journal} {\bibinfo  {journal} {Phys. Scr.}\ }\textbf {\bibinfo {volume} {2005}},\ \bibinfo {pages} {85--88} (\bibinfo {year} {2005})}\BibitemShut {NoStop}%
\bibitem [{\citenamefont {Pati}\ \emph {et~al.}(2007)\citenamefont {Pati}, \citenamefont {Salit}, \citenamefont {Salit},\ and\ \citenamefont {Shahriar}}]{Pati07PRL}%
  \BibitemOpen
  \bibfield  {author} {\bibinfo {author} {\bibfnamefont {G.~S.}\ \bibnamefont {Pati}}, \bibinfo {author} {\bibfnamefont {M.}~\bibnamefont {Salit}}, \bibinfo {author} {\bibfnamefont {K.}~\bibnamefont {Salit}}, \ and\ \bibinfo {author} {\bibfnamefont {M.~S.}\ \bibnamefont {Shahriar}},\ }\bibfield  {title} {\enquote {\bibinfo {title} {Demonstration of a tunable-bandwidth white-light interferometer using anomalous dispersion in atomic vapor},}\ }\href@noop {} {\bibfield  {journal} {\bibinfo  {journal} {Phys. Rev. Lett.}\ }\textbf {\bibinfo {volume} {99}},\ \bibinfo {pages} {133601} (\bibinfo {year} {2007})}\BibitemShut {NoStop}%
\bibitem [{\citenamefont {Wu}\ and\ \citenamefont {Xiao}(2008)}]{Wu08PRA}%
  \BibitemOpen
  \bibfield  {author} {\bibinfo {author} {\bibfnamefont {H.}~\bibnamefont {Wu}}\ and\ \bibinfo {author} {\bibfnamefont {M.}~\bibnamefont {Xiao}},\ }\bibfield  {title} {\enquote {\bibinfo {title} {White-light cavity with competing linear and nonlinear dispersions},}\ }\href@noop {} {\bibfield  {journal} {\bibinfo  {journal} {Phys. Rev. A}\ }\textbf {\bibinfo {volume} {77}},\ \bibinfo {pages} {031801(R)} (\bibinfo {year} {2008})}\BibitemShut {NoStop}%
\bibitem [{\citenamefont {Yum}\ \emph {et~al.}(2013{\natexlab{a}})\citenamefont {Yum}, \citenamefont {Sheuer}, \citenamefont {Salit}, \citenamefont {Hemmer},\ and\ \citenamefont {Shahriar}}]{Yum13JLT}%
  \BibitemOpen
  \bibfield  {author} {\bibinfo {author} {\bibfnamefont {H.~N.}\ \bibnamefont {Yum}}, \bibinfo {author} {\bibfnamefont {J.}~\bibnamefont {Sheuer}}, \bibinfo {author} {\bibfnamefont {M.}~\bibnamefont {Salit}}, \bibinfo {author} {\bibfnamefont {P.~R.}\ \bibnamefont {Hemmer}}, \ and\ \bibinfo {author} {\bibfnamefont {M.~S.}\ \bibnamefont {Shahriar}},\ }\bibfield  {title} {\enquote {\bibinfo {title} {Demonstration of white light cavity effect using stimulated {B}rillouin scattering in a fiber loop},}\ }\href@noop {} {\bibfield  {journal} {\bibinfo  {journal} {J. Lightwave Technol.}\ }\textbf {\bibinfo {volume} {32}},\ \bibinfo {pages} {3865--3872} (\bibinfo {year} {2013}{\natexlab{a}})}\BibitemShut {NoStop}%
\bibitem [{\citenamefont {Wise}\ \emph {et~al.}(2004)\citenamefont {Wise}, \citenamefont {Mueller}, \citenamefont {Reitze}, \citenamefont {Tanner},\ and\ \citenamefont {Whiting}}]{Wise04CQG}%
  \BibitemOpen
  \bibfield  {author} {\bibinfo {author} {\bibfnamefont {S.}~\bibnamefont {Wise}}, \bibinfo {author} {\bibfnamefont {G.}~\bibnamefont {Mueller}}, \bibinfo {author} {\bibfnamefont {D.}~\bibnamefont {Reitze}}, \bibinfo {author} {\bibfnamefont {D.~B.}\ \bibnamefont {Tanner}}, \ and\ \bibinfo {author} {\bibfnamefont {B.~F.}\ \bibnamefont {Whiting}},\ }\bibfield  {title} {\enquote {\bibinfo {title} {Linewidth-broadened {F}abry {P}erot cavities within future gravitational wave detectors},}\ }\href@noop {} {\bibfield  {journal} {\bibinfo  {journal} {Class. Quantum Grav.}\ }\textbf {\bibinfo {volume} {21}},\ \bibinfo {pages} {S1031--S1036} (\bibinfo {year} {2004})}\BibitemShut {NoStop}%
\bibitem [{\citenamefont {Wise}\ \emph {et~al.}(2005)\citenamefont {Wise}, \citenamefont {Quetschke}, \citenamefont {Deshpande}, \citenamefont {Mueller}, \citenamefont {Reitze}, \citenamefont {Tanner}, \citenamefont {Whiting}, \citenamefont {Chen}, \citenamefont {T{\"u}nnermann}, \citenamefont {Kley},\ and\ \citenamefont {Clausnitzer}}]{Wise05PRL}%
  \BibitemOpen
  \bibfield  {author} {\bibinfo {author} {\bibfnamefont {S.}~\bibnamefont {Wise}}, \bibinfo {author} {\bibfnamefont {V.}~\bibnamefont {Quetschke}}, \bibinfo {author} {\bibfnamefont {A.~J.}\ \bibnamefont {Deshpande}}, \bibinfo {author} {\bibfnamefont {G.}~\bibnamefont {Mueller}}, \bibinfo {author} {\bibfnamefont {D.~H.}\ \bibnamefont {Reitze}}, \bibinfo {author} {\bibfnamefont {D.~B.}\ \bibnamefont {Tanner}}, \bibinfo {author} {\bibfnamefont {B.~F.}\ \bibnamefont {Whiting}}, \bibinfo {author} {\bibfnamefont {Y.}~\bibnamefont {Chen}}, \bibinfo {author} {\bibfnamefont {A.}~\bibnamefont {T{\"u}nnermann}}, \bibinfo {author} {\bibfnamefont {E.}~\bibnamefont {Kley}}, \ and\ \bibinfo {author} {\bibfnamefont {T.}~\bibnamefont {Clausnitzer}},\ }\bibfield  {title} {\enquote {\bibinfo {title} {Phase effects in the diffraction of light: {B}eyond the grating equation},}\ }\href@noop {} {\bibfield  {journal} {\bibinfo  {journal} {Phys. Rev. Lett.}\ }\textbf {\bibinfo {volume} {95}},\ \bibinfo {pages} {013901} (\bibinfo
  {year} {2005})}\BibitemShut {NoStop}%
\bibitem [{\citenamefont {Savchenkov}\ \emph {et~al.}(2006)\citenamefont {Savchenkov}, \citenamefont {Matsko},\ and\ \citenamefont {Maleki}}]{Savchenkov06OL}%
  \BibitemOpen
  \bibfield  {author} {\bibinfo {author} {\bibfnamefont {A.~A.}\ \bibnamefont {Savchenkov}}, \bibinfo {author} {\bibfnamefont {A.~B.}\ \bibnamefont {Matsko}}, \ and\ \bibinfo {author} {\bibfnamefont {L.}~\bibnamefont {Maleki}},\ }\bibfield  {title} {\enquote {\bibinfo {title} {White-light whispering gallery mode resonators},}\ }\href@noop {} {\bibfield  {journal} {\bibinfo  {journal} {Opt. Lett.}\ }\textbf {\bibinfo {volume} {31}},\ \bibinfo {pages} {92--94} (\bibinfo {year} {2006})}\BibitemShut {NoStop}%
\bibitem [{\citenamefont {Yum}\ \emph {et~al.}(2013{\natexlab{b}})\citenamefont {Yum}, \citenamefont {Liu}, \citenamefont {Hemmer}, \citenamefont {Scheuer},\ and\ \citenamefont {Shahriar}}]{Yum13OC}%
  \BibitemOpen
  \bibfield  {author} {\bibinfo {author} {\bibfnamefont {H.~N.}\ \bibnamefont {Yum}}, \bibinfo {author} {\bibfnamefont {X.}~\bibnamefont {Liu}}, \bibinfo {author} {\bibfnamefont {P.~R.}\ \bibnamefont {Hemmer}}, \bibinfo {author} {\bibfnamefont {J.}~\bibnamefont {Scheuer}}, \ and\ \bibinfo {author} {\bibfnamefont {M.~S.}\ \bibnamefont {Shahriar}},\ }\bibfield  {title} {\enquote {\bibinfo {title} {The fundamental limitations on the practical realizations of white light cavities},}\ }\href@noop {} {\bibfield  {journal} {\bibinfo  {journal} {Opt. Commun.}\ }\textbf {\bibinfo {volume} {305}},\ \bibinfo {pages} {260--266} (\bibinfo {year} {2013}{\natexlab{b}})}\BibitemShut {NoStop}%
\bibitem [{\citenamefont {Kotlicki}\ and\ \citenamefont {Scheuer}(2014)}]{Kotlicki14OL}%
  \BibitemOpen
  \bibfield  {author} {\bibinfo {author} {\bibfnamefont {O.}~\bibnamefont {Kotlicki}}\ and\ \bibinfo {author} {\bibfnamefont {J.}~\bibnamefont {Scheuer}},\ }\bibfield  {title} {\enquote {\bibinfo {title} {Wideband coherent perfect absorber based on white-light cavity},}\ }\href@noop {} {\bibfield  {journal} {\bibinfo  {journal} {Opt. Lett.}\ }\textbf {\bibinfo {volume} {39}},\ \bibinfo {pages} {6624--6627} (\bibinfo {year} {2014})}\BibitemShut {NoStop}%
\bibitem [{\citenamefont {Shen}\ \emph {et~al.}(2023)\citenamefont {Shen}, \citenamefont {Zhan}, \citenamefont {Wright}, \citenamefont {Christodoulides}, \citenamefont {Wise}, \citenamefont {Willner}, \citenamefont {Zhao}, \citenamefont {Zou}, \citenamefont {Liao}, \citenamefont {Hern{\'a}ndez-Garc{\'i}a}, \citenamefont {Murnane}, \citenamefont {Porras}, \citenamefont {Chong}, \citenamefont {Wan}, \citenamefont {Bliokh}, \citenamefont {Yessenov}, \citenamefont {Abouraddy}, \citenamefont {Wong}, \citenamefont {Go}, \citenamefont {Kumar}, \citenamefont {Guo}, \citenamefont {Fan}, \citenamefont {Papasimakis}, \citenamefont {Zheludev}, \citenamefont {Chen}, \citenamefont {Zhu}, \citenamefont {Agrawal}, \citenamefont {Jolly}, \citenamefont {Dorrer}, \citenamefont {Alonso}, \citenamefont {Lopez-Quintas}, \citenamefont {L{\'o}pez-Ripa}, \citenamefont {Sola}, \citenamefont {Fang}, \citenamefont {Gong}, \citenamefont {Liu}, \citenamefont {Huang}, \citenamefont {Zhang}, \citenamefont {Ruan}, \citenamefont {Mounaix},
  \citenamefont {Fontaine}, \citenamefont {Carpenter}, \citenamefont {Dorrah}, \citenamefont {Capasso},\ and\ \citenamefont {Forbes}}]{Shen23JO}%
  \BibitemOpen
  \bibfield  {author} {\bibinfo {author} {\bibfnamefont {Y.}~\bibnamefont {Shen}}, \bibinfo {author} {\bibfnamefont {Q.}~\bibnamefont {Zhan}}, \bibinfo {author} {\bibfnamefont {L.~G.}\ \bibnamefont {Wright}}, \bibinfo {author} {\bibfnamefont {D.~N.}\ \bibnamefont {Christodoulides}}, \bibinfo {author} {\bibfnamefont {F.~W.}\ \bibnamefont {Wise}}, \bibinfo {author} {\bibfnamefont {A.~E.}\ \bibnamefont {Willner}}, \bibinfo {author} {\bibfnamefont {Z.}~\bibnamefont {Zhao}}, \bibinfo {author} {\bibfnamefont {K.}~\bibnamefont {Zou}}, \bibinfo {author} {\bibfnamefont {C.-T.}\ \bibnamefont {Liao}}, \bibinfo {author} {\bibfnamefont {C.}~\bibnamefont {Hern{\'a}ndez-Garc{\'i}a}}, \bibinfo {author} {\bibfnamefont {M.}~\bibnamefont {Murnane}}, \bibinfo {author} {\bibfnamefont {M.~A.}\ \bibnamefont {Porras}}, \bibinfo {author} {\bibfnamefont {A.}~\bibnamefont {Chong}}, \bibinfo {author} {\bibfnamefont {C.}~\bibnamefont {Wan}}, \bibinfo {author} {\bibfnamefont {K.~Y.}\ \bibnamefont {Bliokh}}, \bibinfo {author}
  {\bibfnamefont {M.}~\bibnamefont {Yessenov}}, \bibinfo {author} {\bibfnamefont {A.~F.}\ \bibnamefont {Abouraddy}}, \bibinfo {author} {\bibfnamefont {L.~J.}\ \bibnamefont {Wong}}, \bibinfo {author} {\bibfnamefont {M.}~\bibnamefont {Go}}, \bibinfo {author} {\bibfnamefont {S.}~\bibnamefont {Kumar}}, \bibinfo {author} {\bibfnamefont {C.}~\bibnamefont {Guo}}, \bibinfo {author} {\bibfnamefont {S.}~\bibnamefont {Fan}}, \bibinfo {author} {\bibfnamefont {N.}~\bibnamefont {Papasimakis}}, \bibinfo {author} {\bibfnamefont {N.~I.}\ \bibnamefont {Zheludev}}, \bibinfo {author} {\bibfnamefont {L.}~\bibnamefont {Chen}}, \bibinfo {author} {\bibfnamefont {W.}~\bibnamefont {Zhu}}, \bibinfo {author} {\bibfnamefont {A.}~\bibnamefont {Agrawal}}, \bibinfo {author} {\bibfnamefont {S.~W.}\ \bibnamefont {Jolly}}, \bibinfo {author} {\bibfnamefont {C.}~\bibnamefont {Dorrer}}, \bibinfo {author} {\bibfnamefont {B.}~\bibnamefont {Alonso}}, \bibinfo {author} {\bibfnamefont {I.}~\bibnamefont {Lopez-Quintas}}, \bibinfo {author}
  {\bibfnamefont {M.}~\bibnamefont {L{\'o}pez-Ripa}}, \bibinfo {author} {\bibfnamefont {{\'I}.~J.}\ \bibnamefont {Sola}}, \bibinfo {author} {\bibfnamefont {Y.}~\bibnamefont {Fang}}, \bibinfo {author} {\bibfnamefont {Q.}~\bibnamefont {Gong}}, \bibinfo {author} {\bibfnamefont {Y.}~\bibnamefont {Liu}}, \bibinfo {author} {\bibfnamefont {J.}~\bibnamefont {Huang}}, \bibinfo {author} {\bibfnamefont {H.}~\bibnamefont {Zhang}}, \bibinfo {author} {\bibfnamefont {Z.}~\bibnamefont {Ruan}}, \bibinfo {author} {\bibfnamefont {M.}~\bibnamefont {Mounaix}}, \bibinfo {author} {\bibfnamefont {N.~K.}\ \bibnamefont {Fontaine}}, \bibinfo {author} {\bibfnamefont {J.}~\bibnamefont {Carpenter}}, \bibinfo {author} {\bibfnamefont {A.~H.}\ \bibnamefont {Dorrah}}, \bibinfo {author} {\bibfnamefont {F.}~\bibnamefont {Capasso}}, \ and\ \bibinfo {author} {\bibfnamefont {A.}~\bibnamefont {Forbes}},\ }\bibfield  {title} {\enquote {\bibinfo {title} {Roadmap on spatiotemporal light fields},}\ }\href@noop {} {\bibfield  {journal} {\bibinfo
  {journal} {J. Opt.}\ }\textbf {\bibinfo {volume} {25}},\ \bibinfo {pages} {093001} (\bibinfo {year} {2023})}\BibitemShut {NoStop}%
\bibitem [{\citenamefont {Abouraddy}\ \emph {et~al.}(2025)\citenamefont {Abouraddy}, \citenamefont {Yessenov}, \citenamefont {Divliansky},\ and\ \citenamefont {Watnik}}]{Abouraddy25OPN}%
  \BibitemOpen
  \bibfield  {author} {\bibinfo {author} {\bibfnamefont {A.~F.}\ \bibnamefont {Abouraddy}}, \bibinfo {author} {\bibfnamefont {M.}~\bibnamefont {Yessenov}}, \bibinfo {author} {\bibfnamefont {I.}~\bibnamefont {Divliansky}}, \ and\ \bibinfo {author} {\bibfnamefont {A.~T.}\ \bibnamefont {Watnik}},\ }\bibfield  {title} {\enquote {\bibinfo {title} {New frontiers in spatiotemporally structured light},}\ }\href@noop {} {\bibfield  {journal} {\bibinfo  {journal} {Opt. Photon. News}\ }\textbf {\bibinfo {volume} {36}},\ \bibinfo {pages} {38--45} (\bibinfo {year} {2025})}\BibitemShut {NoStop}%
\bibitem [{\citenamefont {Torres}\ \emph {et~al.}(2010)\citenamefont {Torres}, \citenamefont {Hendrych},\ and\ \citenamefont {Valencia}}]{Torres10AOP}%
  \BibitemOpen
  \bibfield  {author} {\bibinfo {author} {\bibfnamefont {J.~P.}\ \bibnamefont {Torres}}, \bibinfo {author} {\bibfnamefont {M.}~\bibnamefont {Hendrych}}, \ and\ \bibinfo {author} {\bibfnamefont {A.}~\bibnamefont {Valencia}},\ }\bibfield  {title} {\enquote {\bibinfo {title} {Angular dispersion: an enabling tool in nonlinear and quantum optics},}\ }\href@noop {} {\bibfield  {journal} {\bibinfo  {journal} {Adv. Opt. Photon.}\ }\textbf {\bibinfo {volume} {2}},\ \bibinfo {pages} {319--369} (\bibinfo {year} {2010})}\BibitemShut {NoStop}%
\bibitem [{\citenamefont {Yessenov}\ \emph {et~al.}(2022{\natexlab{a}})\citenamefont {Yessenov}, \citenamefont {Hall}, \citenamefont {Schepler},\ and\ \citenamefont {Abouraddy}}]{Yessenov22AOP}%
  \BibitemOpen
  \bibfield  {author} {\bibinfo {author} {\bibfnamefont {M.}~\bibnamefont {Yessenov}}, \bibinfo {author} {\bibfnamefont {L.~A.}\ \bibnamefont {Hall}}, \bibinfo {author} {\bibfnamefont {K.~L.}\ \bibnamefont {Schepler}}, \ and\ \bibinfo {author} {\bibfnamefont {A.~F.}\ \bibnamefont {Abouraddy}},\ }\bibfield  {title} {\enquote {\bibinfo {title} {Space-time wave packets},}\ }\href@noop {} {\bibfield  {journal} {\bibinfo  {journal} {Adv. Opt. Photon.}\ }\textbf {\bibinfo {volume} {14}},\ \bibinfo {pages} {455--570} (\bibinfo {year} {2022}{\natexlab{a}})}\BibitemShut {NoStop}%
\bibitem [{\citenamefont {Shabahang}\ \emph {et~al.}(2017)\citenamefont {Shabahang}, \citenamefont {Kondakci}, \citenamefont {Villinger}, \citenamefont {Perlstein}, \citenamefont {{El H}alawany},\ and\ \citenamefont {Abouraddy}}]{Shabahang17SR}%
  \BibitemOpen
  \bibfield  {author} {\bibinfo {author} {\bibfnamefont {S.}~\bibnamefont {Shabahang}}, \bibinfo {author} {\bibfnamefont {H.~E.}\ \bibnamefont {Kondakci}}, \bibinfo {author} {\bibfnamefont {M.~L.}\ \bibnamefont {Villinger}}, \bibinfo {author} {\bibfnamefont {J.~D.}\ \bibnamefont {Perlstein}}, \bibinfo {author} {\bibfnamefont {A.}~\bibnamefont {{El H}alawany}}, \ and\ \bibinfo {author} {\bibfnamefont {A.~F.}\ \bibnamefont {Abouraddy}},\ }\bibfield  {title} {\enquote {\bibinfo {title} {Omni-resonant optical micro-cavity},}\ }\href@noop {} {\bibfield  {journal} {\bibinfo  {journal} {Sci. Rep.}\ }\textbf {\bibinfo {volume} {7}},\ \bibinfo {pages} {10336} (\bibinfo {year} {2017})}\BibitemShut {NoStop}%
\bibitem [{\citenamefont {Shabahang}\ \emph {et~al.}(2019)\citenamefont {Shabahang}, \citenamefont {Jahromi}, \citenamefont {Shiri}, \citenamefont {Schepler},\ and\ \citenamefont {Abouraddy}}]{Shabahang19OL}%
  \BibitemOpen
  \bibfield  {author} {\bibinfo {author} {\bibfnamefont {S.}~\bibnamefont {Shabahang}}, \bibinfo {author} {\bibfnamefont {A.~K.}\ \bibnamefont {Jahromi}}, \bibinfo {author} {\bibfnamefont {A.}~\bibnamefont {Shiri}}, \bibinfo {author} {\bibfnamefont {K.~L.}\ \bibnamefont {Schepler}}, \ and\ \bibinfo {author} {\bibfnamefont {A.~F.}\ \bibnamefont {Abouraddy}},\ }\bibfield  {title} {\enquote {\bibinfo {title} {Toggling between active and passive imaging with an omni-resonant micro-cavity},}\ }\href@noop {} {\bibfield  {journal} {\bibinfo  {journal} {Opt. Lett.}\ }\textbf {\bibinfo {volume} {44}},\ \bibinfo {pages} {1532--1535} (\bibinfo {year} {2019})}\BibitemShut {NoStop}%
\bibitem [{\citenamefont {Shiri}\ \emph {et~al.}(2020{\natexlab{a}})\citenamefont {Shiri}, \citenamefont {Yessenov}, \citenamefont {Aravindakshan},\ and\ \citenamefont {Abouraddy}}]{Shiri20OL}%
  \BibitemOpen
  \bibfield  {author} {\bibinfo {author} {\bibfnamefont {A.}~\bibnamefont {Shiri}}, \bibinfo {author} {\bibfnamefont {M.}~\bibnamefont {Yessenov}}, \bibinfo {author} {\bibfnamefont {R.}~\bibnamefont {Aravindakshan}}, \ and\ \bibinfo {author} {\bibfnamefont {A.~F.}\ \bibnamefont {Abouraddy}},\ }\bibfield  {title} {\enquote {\bibinfo {title} {Omni-resonant space-time wave packets},}\ }\href@noop {} {\bibfield  {journal} {\bibinfo  {journal} {Opt. Lett.}\ }\textbf {\bibinfo {volume} {45}},\ \bibinfo {pages} {1774--1777} (\bibinfo {year} {2020}{\natexlab{a}})}\BibitemShut {NoStop}%
\bibitem [{\citenamefont {Shiri}\ \emph {et~al.}(2020{\natexlab{b}})\citenamefont {Shiri}, \citenamefont {Schepler},\ and\ \citenamefont {Abouraddy}}]{Shiri20APLP}%
  \BibitemOpen
  \bibfield  {author} {\bibinfo {author} {\bibfnamefont {A.}~\bibnamefont {Shiri}}, \bibinfo {author} {\bibfnamefont {K.~L.}\ \bibnamefont {Schepler}}, \ and\ \bibinfo {author} {\bibfnamefont {A.~F.}\ \bibnamefont {Abouraddy}},\ }\bibfield  {title} {\enquote {\bibinfo {title} {Programmable omni-resonance using space-time fields},}\ }\href@noop {} {\bibfield  {journal} {\bibinfo  {journal} {APL Photon.}\ }\textbf {\bibinfo {volume} {5}},\ \bibinfo {pages} {106107} (\bibinfo {year} {2020}{\natexlab{b}})}\BibitemShut {NoStop}%
\bibitem [{\citenamefont {Shiri}\ and\ \citenamefont {Abouraddy}(2022)}]{Shiri22OL}%
  \BibitemOpen
  \bibfield  {author} {\bibinfo {author} {\bibfnamefont {A.}~\bibnamefont {Shiri}}\ and\ \bibinfo {author} {\bibfnamefont {A.~F.}\ \bibnamefont {Abouraddy}},\ }\bibfield  {title} {\enquote {\bibinfo {title} {Spatial resolution of omni-resonant imaging},}\ }\href@noop {} {\bibfield  {journal} {\bibinfo  {journal} {Opt. Lett.}\ }\textbf {\bibinfo {volume} {47}},\ \bibinfo {pages} {3804--3807} (\bibinfo {year} {2022})}\BibitemShut {NoStop}%
\bibitem [{\citenamefont {Hall}\ \emph {et~al.}(2025)\citenamefont {Hall}, \citenamefont {Shiri},\ and\ \citenamefont {Abouraddy}}]{Hall25LPR}%
  \BibitemOpen
  \bibfield  {author} {\bibinfo {author} {\bibfnamefont {L.~A.}\ \bibnamefont {Hall}}, \bibinfo {author} {\bibfnamefont {A.}~\bibnamefont {Shiri}}, \ and\ \bibinfo {author} {\bibfnamefont {A.~F.}\ \bibnamefont {Abouraddy}},\ }\bibfield  {title} {\enquote {\bibinfo {title} {Omni-resonant imaging across the visible},}\ }\href@noop {} {\bibfield  {journal} {\bibinfo  {journal} {Laser Photon. Rev.}\ }\textbf {\bibinfo {volume} {19}},\ \bibinfo {pages} {2500270} (\bibinfo {year} {2025})}\BibitemShut {NoStop}%
\bibitem [{\citenamefont {Villinger}\ \emph {et~al.}(2021)\citenamefont {Villinger}, \citenamefont {Shiri}, \citenamefont {Shabahang}, \citenamefont {Jahromi}, \citenamefont {Nasr}, \citenamefont {Villinger},\ and\ \citenamefont {Abouraddy}}]{Villinger21AOM}%
  \BibitemOpen
  \bibfield  {author} {\bibinfo {author} {\bibfnamefont {M.~L.}\ \bibnamefont {Villinger}}, \bibinfo {author} {\bibfnamefont {A.}~\bibnamefont {Shiri}}, \bibinfo {author} {\bibfnamefont {S}~\bibnamefont {Shabahang}}, \bibinfo {author} {\bibfnamefont {A.~K.}\ \bibnamefont {Jahromi}}, \bibinfo {author} {\bibfnamefont {M.~B.}\ \bibnamefont {Nasr}}, \bibinfo {author} {\bibfnamefont {C.}~\bibnamefont {Villinger}}, \ and\ \bibinfo {author} {\bibfnamefont {A.~F.}\ \bibnamefont {Abouraddy}},\ }\bibfield  {title} {\enquote {\bibinfo {title} {Doubling the near-infrared photocurrent in a solar cell via omni-resonant coherent perfect absorption},}\ }\href@noop {} {\bibfield  {journal} {\bibinfo  {journal} {Adv. Opt. Mat.}\ }\textbf {\bibinfo {volume} {9}},\ \bibinfo {pages} {2001107} (\bibinfo {year} {2021})}\BibitemShut {NoStop}%
\bibitem [{\citenamefont {Bhaduri}\ \emph {et~al.}(2020)\citenamefont {Bhaduri}, \citenamefont {Yessenov},\ and\ \citenamefont {Abouraddy}}]{Bhaduri20NatPhot}%
  \BibitemOpen
  \bibfield  {author} {\bibinfo {author} {\bibfnamefont {B.}~\bibnamefont {Bhaduri}}, \bibinfo {author} {\bibfnamefont {M.}~\bibnamefont {Yessenov}}, \ and\ \bibinfo {author} {\bibfnamefont {A.~F}\ \bibnamefont {Abouraddy}},\ }\bibfield  {title} {\enquote {\bibinfo {title} {Anomalous refraction of optical spacetime wave packets},}\ }\href@noop {} {\bibfield  {journal} {\bibinfo  {journal} {Nat. Photon.}\ }\textbf {\bibinfo {volume} {14}},\ \bibinfo {pages} {416--421} (\bibinfo {year} {2020})}\BibitemShut {NoStop}%
\bibitem [{\citenamefont {Benabid}\ \emph {et~al.}(2002)\citenamefont {Benabid}, \citenamefont {Knight}, \citenamefont {Antonopoulos},\ and\ \citenamefont {Russell}}]{Benabid02Science}%
  \BibitemOpen
  \bibfield  {author} {\bibinfo {author} {\bibfnamefont {F.}~\bibnamefont {Benabid}}, \bibinfo {author} {\bibfnamefont {J.~C.}\ \bibnamefont {Knight}}, \bibinfo {author} {\bibfnamefont {G.}~\bibnamefont {Antonopoulos}}, \ and\ \bibinfo {author} {\bibfnamefont {P.~St.~J.}\ \bibnamefont {Russell}},\ }\bibfield  {title} {\enquote {\bibinfo {title} {Stimulated {R}aman scattering in hydrogen-filled hollow-core photonic crystal fiber},}\ }\href@noop {} {\bibfield  {journal} {\bibinfo  {journal} {Science}\ }\textbf {\bibinfo {volume} {298}},\ \bibinfo {pages} {399--402} (\bibinfo {year} {2002})}\BibitemShut {NoStop}%
\bibitem [{\citenamefont {Kondakci}\ and\ \citenamefont {Abouraddy}(2017)}]{Kondakci17NP}%
  \BibitemOpen
  \bibfield  {author} {\bibinfo {author} {\bibfnamefont {H.~E.}\ \bibnamefont {Kondakci}}\ and\ \bibinfo {author} {\bibfnamefont {A.~F.}\ \bibnamefont {Abouraddy}},\ }\bibfield  {title} {\enquote {\bibinfo {title} {Diffraction-free space-time beams},}\ }\href@noop {} {\bibfield  {journal} {\bibinfo  {journal} {Nat. Photon.}\ }\textbf {\bibinfo {volume} {11}},\ \bibinfo {pages} {733--740} (\bibinfo {year} {2017})}\BibitemShut {NoStop}%
\bibitem [{\citenamefont {Hall}\ and\ \citenamefont {Abouraddy}(2024)}]{Hall24OE}%
  \BibitemOpen
  \bibfield  {author} {\bibinfo {author} {\bibfnamefont {L.~A.}\ \bibnamefont {Hall}}\ and\ \bibinfo {author} {\bibfnamefont {A.~F.}\ \bibnamefont {Abouraddy}},\ }\bibfield  {title} {\enquote {\bibinfo {title} {Universal angular-dispersion synthesizer},}\ }\href@noop {} {\bibfield  {journal} {\bibinfo  {journal} {J. Opt. Soc. Am. A}\ }\textbf {\bibinfo {volume} {41}},\ \bibinfo {pages} {83--94} (\bibinfo {year} {2024})}\BibitemShut {NoStop}%
\bibitem [{\citenamefont {Yessenov}\ \emph {et~al.}(2022{\natexlab{b}})\citenamefont {Yessenov}, \citenamefont {Free}, \citenamefont {Chen}, \citenamefont {Johnson}, \citenamefont {Lavery}, \citenamefont {Alonso},\ and\ \citenamefont {Abouraddy}}]{Yessenov22NC}%
  \BibitemOpen
  \bibfield  {author} {\bibinfo {author} {\bibfnamefont {M.}~\bibnamefont {Yessenov}}, \bibinfo {author} {\bibfnamefont {J.}~\bibnamefont {Free}}, \bibinfo {author} {\bibfnamefont {Z.}~\bibnamefont {Chen}}, \bibinfo {author} {\bibfnamefont {E.~G.}\ \bibnamefont {Johnson}}, \bibinfo {author} {\bibfnamefont {M.~P.~J.}\ \bibnamefont {Lavery}}, \bibinfo {author} {\bibfnamefont {M.~A.}\ \bibnamefont {Alonso}}, \ and\ \bibinfo {author} {\bibfnamefont {A.~F.}\ \bibnamefont {Abouraddy}},\ }\bibfield  {title} {\enquote {\bibinfo {title} {Space-time wave packets localized in all dimensions},}\ }\href@noop {} {\bibfield  {journal} {\bibinfo  {journal} {Nat. Commun.}\ }\textbf {\bibinfo {volume} {13}},\ \bibinfo {pages} {4573} (\bibinfo {year} {2022}{\natexlab{b}})}\BibitemShut {NoStop}%
\bibitem [{\citenamefont {Yessenov}\ and\ \citenamefont {Abouraddy}(2025)}]{Yessenov25JOSAA}%
  \BibitemOpen
  \bibfield  {author} {\bibinfo {author} {\bibfnamefont {M.}~\bibnamefont {Yessenov}}\ and\ \bibinfo {author} {\bibfnamefont {A.~F.}\ \bibnamefont {Abouraddy}},\ }\bibfield  {title} {\enquote {\bibinfo {title} {Optical spatiotemporal {F}ourier synthesis: tutorial},}\ }\href@noop {} {\bibfield  {journal} {\bibinfo  {journal} {J. Opt. Soc. Am. A}\ }\textbf {\bibinfo {volume} {42}},\ \bibinfo {pages} {1295--1315} (\bibinfo {year} {2025})}\BibitemShut {NoStop}%
\bibitem [{\citenamefont {Yessenov}\ \emph {et~al.}(2025)\citenamefont {Yessenov}, \citenamefont {Dorrah}, \citenamefont {Guo}, \citenamefont {Hall}, \citenamefont {Park}, \citenamefont {Free}, \citenamefont {Johnson}, \citenamefont {Capasso}, \citenamefont {Fan},\ and\ \citenamefont {Abouraddy}}]{Yessenov25NC}%
  \BibitemOpen
  \bibfield  {author} {\bibinfo {author} {\bibfnamefont {M.}~\bibnamefont {Yessenov}}, \bibinfo {author} {\bibfnamefont {A.~H.}\ \bibnamefont {Dorrah}}, \bibinfo {author} {\bibfnamefont {C.}~\bibnamefont {Guo}}, \bibinfo {author} {\bibfnamefont {L.~A.}\ \bibnamefont {Hall}}, \bibinfo {author} {\bibfnamefont {J.-S.}\ \bibnamefont {Park}}, \bibinfo {author} {\bibfnamefont {J.}~\bibnamefont {Free}}, \bibinfo {author} {\bibfnamefont {E.~G.}\ \bibnamefont {Johnson}}, \bibinfo {author} {\bibfnamefont {F.}~\bibnamefont {Capasso}}, \bibinfo {author} {\bibfnamefont {S.}~\bibnamefont {Fan}}, \ and\ \bibinfo {author} {\bibfnamefont {A.~F.}\ \bibnamefont {Abouraddy}},\ }\bibfield  {title} {\enquote {\bibinfo {title} {Ultrafast space-time optical merons in momentum-energy space},}\ }\href@noop {} {\bibfield  {journal} {\bibinfo  {journal} {Nat. Commun.}\ }\textbf {\bibinfo {volume} {16}},\ \bibinfo {pages} {8592} (\bibinfo {year} {2025})}\BibitemShut {NoStop}%
\bibitem [{\citenamefont {Hall}\ and\ \citenamefont {Abouraddy}(2023)}]{Hall23LPR}%
  \BibitemOpen
  \bibfield  {author} {\bibinfo {author} {\bibfnamefont {L.~A.}\ \bibnamefont {Hall}}\ and\ \bibinfo {author} {\bibfnamefont {A.~F.}\ \bibnamefont {Abouraddy}},\ }\bibfield  {title} {\enquote {\bibinfo {title} {Canceling and inverting normal and anomalous group-velocity dispersion using space-time wave packets},}\ }\href@noop {} {\bibfield  {journal} {\bibinfo  {journal} {Laser Photon. Rev.}\ }\textbf {\bibinfo {volume} {17}},\ \bibinfo {pages} {2200119} (\bibinfo {year} {2023})}\BibitemShut {NoStop}%
\bibitem [{\citenamefont {Yessenov}\ \emph {et~al.}(2023)\citenamefont {Yessenov}, \citenamefont {Mhibik}, \citenamefont {Mach}, \citenamefont {Hayward}, \citenamefont {Menon}, \citenamefont {Glebov}, \citenamefont {Divliansky},\ and\ \citenamefont {Abouraddy}}]{Yessenov23OL}%
  \BibitemOpen
  \bibfield  {author} {\bibinfo {author} {\bibfnamefont {M.}~\bibnamefont {Yessenov}}, \bibinfo {author} {\bibfnamefont {O.}~\bibnamefont {Mhibik}}, \bibinfo {author} {\bibfnamefont {L.}~\bibnamefont {Mach}}, \bibinfo {author} {\bibfnamefont {T.~M.}\ \bibnamefont {Hayward}}, \bibinfo {author} {\bibfnamefont {R.}~\bibnamefont {Menon}}, \bibinfo {author} {\bibfnamefont {L.}~\bibnamefont {Glebov}}, \bibinfo {author} {\bibfnamefont {I.}~\bibnamefont {Divliansky}}, \ and\ \bibinfo {author} {\bibfnamefont {A.~F.}\ \bibnamefont {Abouraddy}},\ }\bibfield  {title} {\enquote {\bibinfo {title} {Ultracompact system for synthesizing space-time wave packets},}\ }\href@noop {} {\bibfield  {journal} {\bibinfo  {journal} {Opt. Lett.}\ }\textbf {\bibinfo {volume} {48}},\ \bibinfo {pages} {2500--2503} (\bibinfo {year} {2023})}\BibitemShut {NoStop}%
\end{thebibliography}%

\clearpage
\vspace{2mm}
\noindent
\textbf{Supplementary}\\
\noindent We present here additional calculations regarding the fraction of incident energy from a conventional focused Gaussian pulse and an omni-resonant space-time wave packet coupled to a planar Fabry-P{\'e}rot cavity. Furthermore, we present calculations of the cavity enhanced intensity for a focused Gaussian pulse normalized to a different reference from that used in the main text.
\\
\\

The fraction of energy from a pulsed laser beam coupled to a planar Fabry-P{\'e}rot (FP) cavity is determined by the overlap between the spatiotemporal spectrum of the incident optical field $|\widetilde{\psi}(k_{x},\omega)|^{2}$ and the spatiotemporal spectral transmission of the FP cavity $T_{\mathrm{FP}}(k_{x},\omega)$, both of which are defined in the main text:
\begin{equation}
\Delta\mathcal{E}=\iint\! dk_{x}d\omega\;|\widetilde{\psi}(k_{x},\omega)|^{2}T_{\mathrm{FP}}(k_{x},\omega);
\label{eq:energy_coupling}
\end{equation}
here $\omega$ is the temporal frequency and $k_{x}$ is the transverse wave number along $x$. In the main text we carried out this calculation for two field configurations: a focused Gaussian pulse for which we obtain $\Delta\mathcal{E}_{\mathrm{G}}$, and an omni-resonant space-time wave packet (STWP) for which we obtain $\Delta\mathcal{E}_{\mathrm{ST}}$.

\begin{figure}[b!]
\centering
\includegraphics[width=8.6cm]{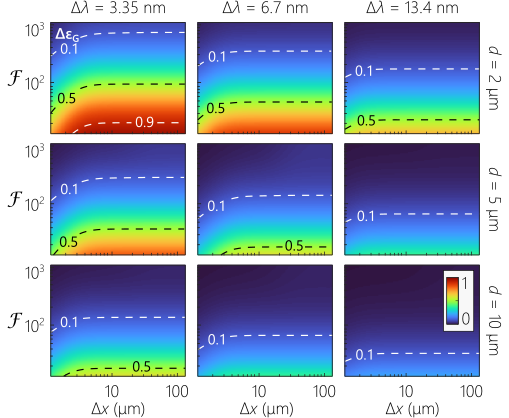}
\caption{Energy fraction $\Delta\mathcal{E}_{\mathrm{G}}$ of a focused Gaussian pulse coupled to a planar FP cavity. Each column corresponds to a fixed temporal bandwidth $\Delta\lambda=3.35,6.7$, and 13.4~nm, and each row corresponds to a different cavity length $d=2,5$, and 10~$\upmu$m. In each panel we vary the spectral uncertainty $\delta\lambda_{\mathrm{ST}}$ and the cavity finesse $\mathcal{F}$. The central panel corresponds to Fig.~2(a) in the main text. The initial energy is held constant throughout.}
\label{fig:EC_Gauss}
\end{figure}

In the main text, we plot $\Delta\mathcal{E}_{\mathrm{G}}$ for a focused Gaussian pulse in Fig.~2(a) for the special case of a temporal bandwidth $\Delta\lambda=6.7$~nm and a cavity length $d=5$~$\upmu$m. Here, we plot $\Delta\mathcal{E}_{\mathrm{G}}$ in Fig.~\ref{fig:EC_Gauss} for three temporal bandwidths $\Delta\lambda=3.35,6.7$, and 13.4~nm, and for cavity lengths $d=2,5$, and 10~$\upmu$m. In each panel in Fig.~\ref{fig:EC_Gauss} (where $\Delta\lambda$ and $d$ are fixed) we vary the focused spot size $\Delta x$ (from 1 to 125~$\upmu$m) and the cavity finesse $\mathcal{F}$ (from 10 to 1000). The central panel in Fig.~\ref{fig:EC_Gauss} here therefore corresponds to Fig.~2(a) in the main text. Within each panel, $\Delta\mathcal{E}_{\mathrm{G}}$ drops with increasing finesse $\mathcal{F}$ (due to the spectral filtering associated with decreasing resonant linewidth $\delta\lambda$), and is approximately independent of the spot size, except at for smaller spot sizes where $\Delta\mathcal{E}_{\mathrm{G}}$ gradually drops. 

The overall structures of $\Delta\mathcal{E}_{\mathrm{G}}$ in the panels are similar, with the main difference being the maximum value reached; i.e., the panels in Fig.~\ref{fig:EC_Gauss} differ from each other approximately with only an overall scaling factor. We observe that $\Delta\mathcal{E}_{\mathrm{G}}$ increases when $d$ decreases because the resonant cavity linewidth $\delta\lambda$ increases, thus reducing spectral filtering. Similarly, $\Delta\mathcal{E}_{\mathrm{G}}$ increases when $\Delta\lambda$ is reduced at fixed $d$, because the impact of spectral filtering is reduced. These results are useful in interpreting Fig.~3 in the main text.

\begin{figure}[t!]
\centering
\includegraphics[width=8.6cm]{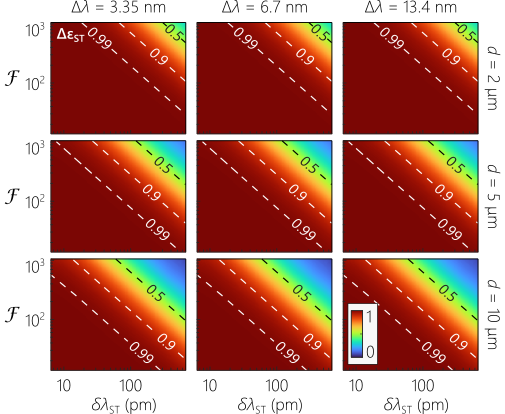}
\caption{Energy fraction $\Delta\mathcal{E}_{\mathrm{ST}}$ of an omni-resonant STWP coupled to a planar FP cavity. Each column corresponds to a fixed temporal bandwidth $\Delta\lambda=3.35,6.7$, and 13.4~nm, and each row corresponds to a different cavity length $d=2,5$, and 10~$\upmu$m. In each panel we vary the spectral uncertainty $\delta\lambda_{\mathrm{ST}}$ and the cavity finesse $\mathcal{F}$. The central panel corresponds to Fig.~2(b) in the main text. The initial energy is held constant throughout.}
\label{fig:EC_omni}
\end{figure}

In Fig.~\ref{fig:EC_omni} we plot $\mathcal{E}_{\mathrm{ST}}$ for omni-resonant space-time wave packets (STWPs) for temporal bandwidths $\Delta\lambda=3.35,6.7$, and $13.4$~nm and $d=2,5$, and 10~$\upmu$m as done in Fig.~\ref{fig:EC_Gauss}. In each panel in Fig.~\ref{fig:EC_omni}, we vary the spectral uncertainty $\delta\lambda_{\mathrm{ST}}$ of the STWP from 6 to 600~pm, and we vary the cavity finesse $\mathcal{F}$ over the range 10 to 1000. The central panel in Fig.~\ref{fig:EC_omni} where $\Delta\lambda=6.7$~nm and $d=5$~$\upmu$m corresponds to Fig.~2(b) in the main text. We find that $\Delta\mathcal{E}_{\mathrm{ST}}$ is almost invariant as we change $\Delta\lambda$ (and thus the pulse width) because the spatiotemporal spectrum of the omni-resonant STWP matches the spatiotemporal spectral transfer function of the FP cavity, which enables the entire spectrum of the pulse to couple independently of its bandwidth. However, when the cavity length $d$ increases, the resonant linewidth $\delta\lambda$ decreases, which increases the impact of spectral filtering on the $\Delta\mathcal{E}_{\mathrm{ST}}$. These results are useful for interpreting Fig.~4 in the main text.

\begin{figure}[t!]
\centering
\includegraphics[width=8.6cm]{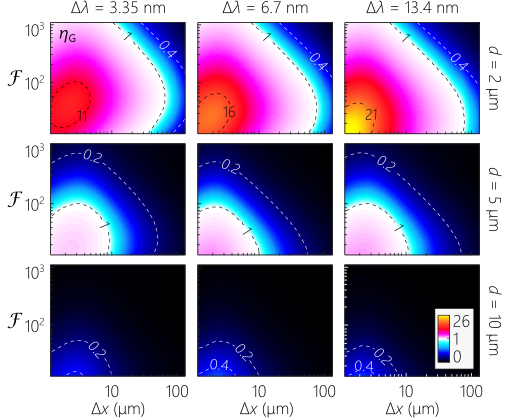}
\caption{Intensity enhancement for a focused Gaussian pulse $\eta_{\mathrm{G}}$. Unlike Fig.~3 in the main manuscript, here the enhancement is calculated with respect to a free-space focused Gaussian pulse with a \textit{constant} spot size in each panel. The spot sizes considered here are 1.8, 1.3, 0.9~$\upmu$m from left to right. These values are the spot sizes associated with the omni-resonant STWPs of equal temporal bandwidth $\Delta\lambda$.}
\label{fig:Gauss_enhancement}
\end{figure}

Finally, we plot in Fig.~\ref{fig:Gauss_enhancement} the enhancement factor $\eta_{\mathrm{G}}=I_{\mathrm{peak}}^{\mathrm{G,FP}}/I_{\mathrm{peak}}^{\mathrm{G,free}}$, which compares the intra-cavity peak intensity for the focused Gaussian pulse with respect to its free-space counterpart -- using the same parameters from Fig.~\ref{fig:EC_Gauss}. This differs from Fig.~3 in the main text in one respect: in Fig.~3 in the main text we normalized each intra-cavity focused Gaussian pulse with a free-space counterpart having the same energy, pulse width, and spot size, whereas is Fig.~\ref{fig:Gauss_enhancement} here we hold the spot size fixed for the free-space focused Gaussian spot in each panel. We select the spot size $\Delta x$ to be the same as that for an omni-resonant STWP having the same temporal bandwith. These spot sizes are 1.8, 1.3, and 0.9~$\upmu$m for the temporal bandwidths $\Delta\lambda=3.35,6.7$, and 13.4~nm. The structure of $\eta_{\mathrm{G}}$ with this normalization scheme differ significantly with that in Fig.~3 in the main text. However, the two share a critical feature: $\eta_{\mathrm{G}}$ drops with increasing cavity length $d$ due to spectral filtering.

\end{document}